%% file: submittedICC2026Final.tex
\documentclass[conference,a4paper,twoside]{IEEEtran}

\IEEEoverridecommandlockouts

\newcommand\hmmax{0}
\newcommand\bmmax{0}

\usepackage{graphicx}
\usepackage{tikz}
\usepackage{pgfplots}
\usepackage{xcolor}
\usepackage{color, colortbl}
\usepackage{amsmath}
\usepackage{bbm}
\usepackage{amsfonts, amssymb}
\usepackage{mathtools}
\usepackage{cite}
\usepackage[nolist]{acronym}

\usepackage{booktabs} 		
\usepackage{diagbox}		
\usepackage{upgreek}
\usepackage{bbm}
\usepackage{Files/bm}
\usepackage{Files/winsnotation}
\usepackage{Files/LVcolours} 

\usepackage[ruled,vlined,noresetcount]{algorithm2e}
\usepackage{setspace}	 	
\usepackage{comment}
\usepackage{multirow}

\usepackage{array}
\newcolumntype{C}[1]{>{\centering\arraybackslash}p{#1}}

\makeatletter
\def\endthebibliography{%
  \def\@noitemerr{\@latex@warning{Empty `thebibliography' environment}}%
  \endlist
}
\makeatother
\usepackage{amsthm}
\theoremstyle{definition}

\DeclareMathOperator*{\argmin}{arg\,min}

\pgfplotsset{compat=1.17}
\usepgfplotslibrary{fillbetween}

\begin{document}

\title{Practical Low-Weight Codes for \\ Energy-Efficient Bus Encoding}

\author{%
  \IEEEauthorblockN{Lorenzo~Valentini,~\IEEEmembership{Member,~IEEE,} Marco~Chiani,~\IEEEmembership{Fellow,~IEEE}}
  \IEEEauthorblockA{CNIT/WiLab, DEI, University of Bologna, Italy \\
  	Email: \{lorenzo.valentini13,  marco.chiani\}@unibo.it }
}


\maketitle 

\input{Files/Acronimi_CD.tex}

\begin{abstract}

We consider the transmission of data encoded into binary messages, with the goal of minimizing the Hamming distance, i.e., the number of bit-flips, between consecutive messages. This problem is relevant for enhancing the longevity of Non-Volatile Memories and reducing transition-induced energy consumption in data buses. 
Known as Write-Efficient Memory coding in the literature, this challenge has traditionally been addressed using optimal but complex schemes.
In low-power computer systems the same topic is known as bus encoding.
In this paper, we derive closed-form expressions to evaluate the average number of bit-flips for practical, sub-optimal encoding schemes, and propose two new schemes assisted by predefined random codebooks. 
We demonstrate that low-complexity solutions achieve performance very close to the optimal schemes, making them attractive for implementation in energy-sensitive and memory-critical applications.
For instance, by adding $8$ extra bits to $64$-bits data, sub-optimal schemes can achieve a bit-flip reduction (related to energy saving) of approximately 24.7\%, compared to the 26.4\% reduction offered by the significantly more complex optimal scheme.

\end{abstract}


\section{Introduction}

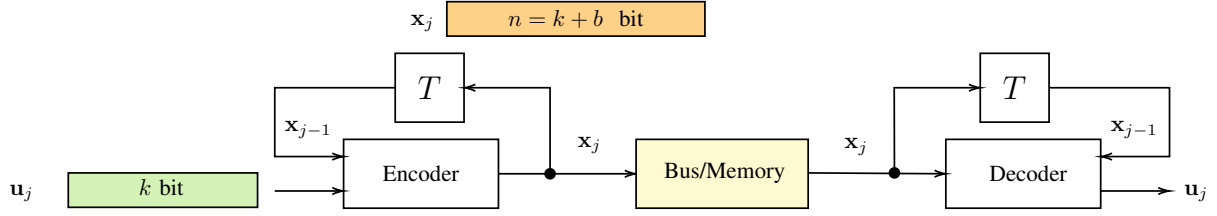
\begin{figure*}[t]
    \centering
    \resizebox{0.9\textwidth}{!}{   \input{Figures/GeneralFrameworkFinal}
    }
    \caption{General Framework. The block with the label $T$ indicates a delay line.}
    \label{fig:GenFrame}
\end{figure*}

For data communication systems, minimizing the Hamming distance between consecutive messages is often critical, as it directly impacts energy consumption, system longevity, and reliability. In coding theory, this scenario is referred to as binary Hamming \ac{WEM} \cite{Ahl:89}. This problem is also connected to steganography, where the goal is to embed information with minimal distortion \cite{BasPin:99centered,GalKab:09stega}. Several practical applications highlight the relevance of this problem. For instance, in \ac{NVM}, the number of permissible write operations directly affects memory lifespan \cite{Lee:10}. Similarly, in low-power computer systems, transitions on data buses contribute significantly to dynamic power consumption  \cite{Jaf:14,Vel:20,CheEtzSch:24}. 
In addition to their conventional application in memory interfaces \cite{Connor22:BusEncodingPAMMemory}, with the increasing number of artificial intelligence applications, the efficient use of buses is also becoming an important challenge in the implementation of \acp{CNN}, where systolic array architectures are used to perform matrix multiplications \cite{peltekis2023low}.
To address this, a variety of bus encoding techniques have been proposed and adopted, including in high-performance GPU memory systems \cite{Ben:00,Connor22:BusEncodingPAMMemory}.
While the underlying technical frameworks are essential, understanding their real-world implications can make the subject more tangible. 
For example, in large-scale data centers, even minor improvements in energy efficiency per data bus operation can scale to substantial energy savings, highlighting the economic and environmental stakes.
Similarly, the impact of these techniques could provide saving in Network-on-Chip applications where it is estimated that the dynamic power dissipated by interconnections is over $50\%$ of the total dynamic power~\cite{Mag:04,Wol:08,Jaf:14,Vel:20}.

In this work, we will use the term ``bus encoding'' as a general label for all techniques aimed at reducing the Hamming distance between consecutive messages.
One of the earliest and simplest examples of such techniques is the \ac{DBI} scheme, patented for bus encoding and widely known in hardware design \cite{Fle:87,Sta:95, Cha24:AppBusEnc}. In \ac{DBI}, $k$ data bits are encoded into an $n = k + 1$ bit message by either transmitting the original data or its bitwise complement, whichever yields a smaller Hamming distance from the previous message. An extra bit signals whether inversion has been applied. This concept has been rediscovered in different contexts, such as in \ac{PRAM} under the name Flip-N-Write, and is recognized in \ac{WEM} theory as a coset code based on repetition coding \cite{Cho:09,Fan:00}.

Theoretical bounds for \ac{WEM} have been established in \cite{Ahl:89,Fan:00}. Specifically, if \(D\) is the average number of bit-flips and \(r = k/n\) is the rate, then for \(n \to \infty\), the fundamental limit is \(r = h(D/n)\), where \(h(\cdot)\) denotes the binary entropy function \cite[Th. 4]{Ahl:89}. In practice, finite values of \(k\), such as \(k = 64\) in bus encoding, require the design of finite-length codes. Assuming uniformly distributed \ac{i.i.d.} $k$-bit data, and defining the redundancy as \(b = n - k\), it is known that \ac{DBI} is optimal for the minimum redundancy case \(b = 1\). 
For \(b > 1\), minimizing bit-flips requires mapping the data into $n$-bit codewords of lowest possible Hamming weight, then transmitting the XOR between the selected codeword and the previous message. This approach has emerged independently across different domains, including in \cite{Tab:90}, where both lookup-table and algorithmic implementations are suggested. More efficient implementations are those based on the combinatorial number system \cite{Dav:66,Cov:73,Ram:99,WuJia:11,ValChi:24arxiv}. From a coding-theoretic perspective, the same principle is equivalent to a coset coding scheme, where the lowest-weight representative of a coset is chosen \cite{Ahl:89,Coh:90,Fan:00,Jac:13}.

However, while optimal, such schemes can be impractical for large $k$ and $b$, due to their computational and memory complexity. This has motivated the development of sub-optimal schemes that offer near-optimal performance with significantly reduced complexity.

\vspace{0.3em}
\noindent The main contributions of this paper are:
\begin{itemize}
    \item Derivation of closed-form expressions for the average bit-flips of several practical sub-optimal bus encoding schemes.
    \item Proposal of two new sub-optimal bus encoding schemes.
    \item Demonstration that some sub-optimal schemes, while simpler to implement, closely approach the performance of optimal schemes.
\end{itemize}

Throughout the paper, we adopt the following notation: boldface letters $\mathbf{x}$ denote binary vectors; $\E{\cdot}$ is the expectation operator; $\Prob{\cdot}$ indicates probability; $d_\mathrm{H}(\mathbf{a},\mathbf{b})$ denotes the Hamming distance; and $w(\mathbf{a})$ is the Hamming weight. Concatenation is written as $\mathbf{a} \, || \, \mathbf{b}$, complement as $\overline{\mathbf{a}}$, and bitwise XOR as $\mathbf{a} \oplus \mathbf{b}$.

\section{Problem Formulation}
\label{sec:preliminary}
In this section, we formally define the problem and review relevant foundational results.
\subsection{Reducing Bit-Flips}\label{subsec:GenFrame}

Consider a \ac{DMS} that emits binary $k$-tuples $\mathbf{u}_1, \mathbf{u}_2, \ldots$, where $\mathbf{u}_j$ denotes the data at time $j$. Such data sequences can represent, for example, words transmitted over a system bus, where minimizing the number of line transitions is desirable, or data to be written on a non-volatile memory, where the goal is to reduce the number of bit-flips relative to the current content.
For example, in bus encoding literature, we can reduce the dynamic power consumption reducing the number of bit-flips~\cite{Jaf:14,Vel:20,CheEtzSch:24}. The source $\mathbf{u}_j$ is assumed to produce independent, uniformly distributed elements in the alphabet $\{0,1\}^k$.
As shown in Fig.~\ref{fig:GenFrame}, each data word $\mathbf{u}_j$ is encoded into a codeword $\mathbf{x}_j \in \{0,1\}^n$, where $n = k + b$ and $b$ denotes the number of added redundancy bits (e.g., additional bus lines). The objective is to minimize the average Hamming distance between consecutive messages, given by
\begin{align}
\label{eq:DGenFrame}
    D(k,b) = \E{d_\mathrm{H}(\mathbf{x}_{j},\,\mathbf{x}_{j-1})} = \E{w(\mathbf{x}_{j} \oplus \mathbf{x}_{j-1})},
\end{align}
where the expectation is taken over all possible data sequences.

In the uncoded case ($b = 0$), we have $\mathbf{x}_j = \mathbf{u}_j$, and the average distance becomes 
$    D_\mathrm{unc}(k) = D(k,0) = {k}/{2}.
$ 
We will evaluate $D(k,b)$ for various encoding schemes and compare their efficiency using the normalized bit-flip savings
\begin{equation}\label{eq:1-D/Dunc}
    \eta = 1 - \frac{D(k,b)}{D_\mathrm{unc}(k)}.
\end{equation}
This metric captures the relative improvement over the uncoded baseline.

\subsection{Optimal $(k,b)$ Scheme and Its Performance}\label{subsec:Bound}

Assume $2^k$ equiprobable data $k$-tuples are to be mapped to $n$-bit codewords. The optimal encoder chooses a codebook $\mathcal{D}$ containing the $2^k$ lowest-weight $n$-tuples. This set includes all $n$-bit words of weight 0, then weight 1, and so on, up to the minimum required to reach cardinality $2^k$. Each input $\mathbf{u}_j$ is mapped to a codeword $\mathbf{d}_j \in \mathcal{D}$, and the transmitted message is formed via differential encoding
\begin{align}
    \mathbf{x}_{j} = \mathbf{d}_{j} \oplus \mathbf{x}_{j-1}
\end{align}
as discussed in \cite{Ahl:89,Tab:90,Sta:95}.

The probability that the message $\mathbf{x}_j$ differs from $\mathbf{x}_{j-1}$ in exactly $d$ bits is given by
\begin{align}
\label{eq:pOpt}
    p_\mathrm{opt}(d) =
    \begin{dcases}
        \binom{n}{d} \, 2^{-k}, & \text{if } 0 \leq d < d_\mathrm{max}, \\
        1 - \sum_{i=0}^{d_\mathrm{max}-1} \binom{n}{i} \, 2^{-k}, & \text{if } d = d_\mathrm{max},
    \end{dcases}
\end{align}
where $d_\mathrm{max}$ is the smallest integer such that $\sum_{i=0}^{d_\mathrm{max}} \binom{n}{i} \geq 2^k$.

Accordingly, the average Hamming distance for the optimal scheme is
\begin{align}
\label{eq:Dopt}
    D_\mathrm{opt}(k, b) = d_\mathrm{max} - \sum_{i = 0}^{d_\mathrm{max} - 1} \frac{d_\mathrm{max} - i}{2^k} \binom{n}{i}.
\end{align}
This expression, derived in \cite[Th. 4]{Ahl:89}, serves as the benchmark for all sub-optimal schemes discussed in this work. Note that the maximum useful redundancy is $b = 2^k - 1 - k$, which permits encoding all data using only weight-one codewords (e.g., via \ac{PPM} with inclusion of the all-zero vector).

\subsubsection{Case $b = 1$: Data Bus Inversion (DBI)}\label{subsec:Singleinv}
When just one redundancy bit is available ($b = 1$), the optimal strategy is known as Data Bus Inversion (\ac{DBI}) \cite{Fle:87,Tab:90,Sta:95}, and is widely used due to its simplicity. 
In \ac{DBI}, either the original data word $\mathbf{u}_j$ is transmitted with a trailing $0$, or its bitwise complement $\overline{\mathbf{u}}_j$ is transmitted with a trailing $1$, depending on which choice yields fewer bit-flips compared to the previous message
\[
    \mathbf{x}_{j} =
    \begin{cases}
        \mathbf{u}_j \, || \, 0, & \text{if } d_\mathrm{H}(\mathbf{u}_j \, || \, 0, \mathbf{x}_{j-1}) \leq (k+1)/2, \\
        \overline{\mathbf{u}}_j \, || \, 1, & \text{otherwise}.
    \end{cases}
\]
The decoder simply checks the final bit to determine whether to apply inversion.
It is simple to check that this is the optimal scheme when just one bit is added. The average distance $D_\mathrm{DBI}(k) \triangleq D_\mathrm{opt}(k,1)$ is given by \eqref{eq:Dopt} with $d_\mathrm{max}=\lceil k/2 \rceil$. The resulting formula for $D_\mathrm{DBI}(k)$  is equivalent to the expressions in \cite{Tab:90,Sta:95}.

\subsubsection{Implementation of the Optimal Scheme}
Implementing the optimal encoder requires a mapping from $\{0,1\}^k$ to the set $\mathcal{D}$ of $2^k$ lowest-weight codewords. For $b = 1$, this mapping is simple, as \ac{DBI} is optimal. For $b > 1$, practical implementations may involve: $i)$ lookup tables \cite{Tab:90}; $ii)$ combinatorial number system indexing \cite{Ram:99}; $iii)$ syndrome decoding using coset codes \cite{Ahl:89,Coh:90,Fan:00,Jac:13}.
However, the increased complexity associated with these approaches may limit their use in practical systems.

In the following sections, we explore sub-optimal schemes that offer simpler implementation while retaining much of the performance advantage of the optimal strategy.

\section{Performance Analysis of Sub-Optimal Schemes}
\label{sec:SubOpt}

This section presents the main novel contributions of the paper. Specifically, we analyze the most relevant practical schemes and introduce two new strategies, Random \& Inversion and Shift \& Inversion, along with their analytical performance evaluation. All expressions in this section are original. To the best of our knowledge, the only prior analytical results available in the literature is equation \eqref{eq:Dopt} for the optimal scheme described in the previous section.

\subsection{Partitioned Inversion Scheme}\label{subsec:Pinv}

We begin by analyzing the average bit-flip distance achieved by the \ac{PI} scheme proposed in \cite{Sta:95}. The idea is to partition the $k$-bit input sequence into $b$ blocks and apply the \ac{DBI} technique independently to each block. Each additional redundancy bit is then used to indicate whether inversion was applied to the corresponding block.
When $k$ is not evenly divisible by $b$, we partition the input as uniformly as possible. For instance, with $k = 16$ and $b = 3$, the block lengths would be $6$, $5$, and $5$.
The average bit-flip distance for the \ac{PI} scheme can then be expressed as
\begin{align}
    D_\mathrm{PI}(k, b) &= \big[b - (k \bmod b)\big] \cdot D_\mathrm{DBI}\left(\left\lfloor \frac{k}{b} \right\rfloor\right) \nonumber \\
    &\quad + (k \bmod b) \cdot D_\mathrm{DBI}\left(\left\lfloor \frac{k}{b} \right\rfloor + 1\right),
\end{align}
where $D_\mathrm{DBI}(\cdot) = D_\mathrm{opt}(\cdot,1)$.

\subsection{Random Scheme}\label{subsec:PureRnd}


Next, we consider the \ac{PR} scheme, which uses a codebook of randomly pre-generated sequences. Since the codebook must be known to both the encoder and the decoder, it can be generated and shared once and subsequently kept fixed. This approach is similar to the method proposed in \cite{Kom:99} for bus encoding and has also been explored in the context of \ac{NVM} in \cite{Sey:16}.

Specifically, the encoder and decoder share a common codebook $\{\mathbf{r}^{(i)}\}_{i = 0}^{L-1}$ consisting of $L = 2^b$ randomly generated binary sequences of length $k$. A more accurate codebook design can be pursued from the codewords of a classical error correcting
code, like a Reed-Muller. For each data word $\mathbf{u}_j$, the encoder computes $L$ candidate messages and selects the one with the smallest Hamming distance to the previous transmitted message $\mathbf{x}_{j-1}$.
The $i$-th candidate is formed by
\begin{equation}\label{eq:xjiPR}
    \mathbf{u}^{(i)}_j = \mathbf{u}_j \oplus \mathbf{r}^{(i)}, \qquad  
    \mathbf{x}_{j}^{(i)} = \mathbf{u}^{(i)}_j \,||\, \mathbf{a}^{(i)},
\end{equation}
where $\mathbf{a}^{(i)}$ is the $b$-bit binary representation of $i$. The transmitted message is then
\begin{equation}\label{eq:xjargmin}
    \mathbf{x}_j = \argmin_{\mathbf{x}_{j}^{(i)}} d_\mathrm{H}(\mathbf{x}_{j}^{(i)}, \mathbf{x}_{j-1})\,.
\end{equation}
The decoder extracts $\mathbf{a}^{(i)}$ and uses it to identify $\mathbf{r}^{(i)}$, then recovers $\mathbf{u}_j = \mathbf{u}^{(i)}_j \oplus \mathbf{r}^{(i)}$.

To analyze the performance, define the random variable
\[
    \rv{Y}^{(w_i)} = d_\mathrm{H}(\mathbf{x}_{j}^{(i)}, \mathbf{x}_{j-1}) = \rv{D}^{(i)} + w_i,
\]
where $\rv{D}^{(i)}$ is the Hamming distance between $\mathbf{u}^{(i)}_j$ and the last $k$ bits of $\mathbf{x}_{j-1}$, and $w_i$ is the distance contribution from the index bits $\mathbf{a}^{(i)}$.
Since $\mathbf{u}^{(i)}_j$ and $\mathbf{x}_{j-1}$ are uniformly random and independent, the distribution of $\rv{D}^{(i)}$ is
\begin{align}
\label{eq:PdistGivenk}
    \Prob{\rv{D}^{(i)} = d} = \binom{k}{d} \cdot 2^{-k}.
\end{align}
For each $w = 0, \dots, b$, there are $\binom{b}{w}$ candidates with index contribution $w_i = w$. Recalling that for a non-negative random variable $\rv{X}$ taking values in $\mathbb{N}$ we have  $\E{\rv{X}} = \sum_{m \ge 0} \Prob{\rv{X} > m}$  \cite{casella2021statistical}, the expected minimum Hamming distance is
\begin{align}
\label{eq:DPR}
    D_\mathrm{PR}(k, b) &= \E{\min_i \rv{Y}^{(w_i)}} \nonumber \\
    &= \sum_{m=0}^{k-1} \prod_{w = 0}^{b} \left[ \sum_{\ell = m - w + 1}^{k} \binom{k}{\ell} \cdot 2^{-k} \right]^{\binom{b}{w}}.
\end{align}

An approximation can be made by modeling all $n = k + b$ bits in $\mathbf{x}_j^{(i)}$ as independent and uniformly random
\begin{align}
\label{eq:DPRApprox}
    D_\mathrm{PR}(k, b) 
    \approx \sum_{m=0}^{n-1} \left[ \sum_{\ell = m+1}^{n} \binom{n}{\ell} \cdot 2^{-n} \right]^L.
\end{align}

\subsection{Random \& Inversion Scheme}\label{subsec:RndInv}

We now introduce a second novel scheme, called \ac{RI}, which extends the \ac{PR} scheme by incorporating bus inversion.
In this scheme, one of the $b$ bits is reserved for bus inversion, following the principles of \ac{DBI}. The remaining $b-1$ bits index $L/2 = 2^{b-1}$ randomly generated sequences $\{\mathbf{r}^{(i)}\}_{i=0}^{L/2 - 1}$, while the other half of the codebook is their complement: $\mathbf{r}^{(i+L/2)} = \overline{\mathbf{r}}^{(i)}$.
The encoder computes $L$ candidate messages
\[
\mathbf{x}_{j}^{(i)} = \mathbf{u}_j \oplus \mathbf{r}^{(i)} \,||\, \mathbf{a}^{(i)}, \quad 
\mathbf{x}_{j}^{(i+L/2)} = \mathbf{u}_j \oplus \overline{\mathbf{r}}^{(i)} \,||\, \mathbf{a}^{(i+L/2)},
\]
and selects the one closest to $\mathbf{x}_{j-1}$, as in \eqref{eq:xjargmin}.

The effective Hamming distance is
\begin{align}
\label{eq:DRIa}
    \rv{Y}^{(w_i)} &= \min\left( \rv{D}^{(i)} + w_i,\; k + 1 - \rv{D}^{(i)} + w_i \right),
\end{align}
where $w_i$ is the weight of the index part $\mathbf{a}^{(i)}$.
Using this, the distribution becomes
\begin{align}
\label{eq:DRIb}
    \Prob{\rv{Y}^{(w_i)} = \ell} =
    \begin{cases}
        \frac{1}{2^k} \binom{k+1}{\ell - w_i}, & \ell < w_i + \frac{k+1}{2}, \\
        \frac{1}{2^k} \binom{k}{\ell - w_i},   & \ell = w_i + \frac{k+1}{2}.
    \end{cases}
\end{align}
Finally, the average minimum distance is
\begin{align}
\label{eq:DRI}
    D_\mathrm{RI}(k, b) &= \sum_{m=0}^{k-1} \prod_{w = 0}^{b-1}
    \left[
        \sum_{\ell = m+1}^{\lfloor w + (k+1)/2 \rfloor} \Prob{\rv{Y}^{(w)} = \ell}
    \right]^{\binom{b-1}{w}}.
\end{align}

\subsection{Shift \& Inversion Scheme}\label{subsec:ShiftInv}

We now introduce and analyze a novel scheme, called Shift \& Inversion (\ac{SI}), which is a low-complexity variant of \ac{RI} where the pre-shared codebook is replaced with circular shifts of the data itself, and combined with bus inversion. Comparing circular shifts of the data with the previous data bus, instead of using a shared codebook, is an effective idea that has been proposed, e.g., in \cite{Ala:15}, but not combined with bus inversion. 
In the \ac{SI} scheme, given a data word $\mathbf{u}_j$, the encoder generates $2^{b-1}$ shifted versions
\[
\mathbf{u}^{(i)}_j = \text{shift}(\mathbf{u}_j, i),
\]
for $i = 0, \ldots, L/2-1$, and uses their complements for the second half. These are combined with index vectors to form candidate messages, as in the RI scheme. The best candidate is selected as in \eqref{eq:xjargmin}.
Unlike PR and RI, this scheme does not require a shared codebook between encoder and decoder. The decoder simply extracts the shift and inversion flags from the index field and reverses them.

Since circular shifts approximate independent variation for moderate-to-large $k$, the SI scheme performs similarly to RI as long as $2^{b-1} \leq k$ (i.e., shifts are non-redundant). Thus, the performance is approximated by
\begin{align}
\label{eq:DSI}
    D_\mathrm{SI}(k, b) \approx
    \begin{dcases}
        D_\mathrm{RI}(k, b), & \text{if } 2^{b-1} \leq k, \\
        D_\mathrm{RI}(k, 1 + \lceil \log_2(k) \rceil), & \text{otherwise}.
    \end{dcases}
\end{align}
This approximation has been verified through simulations, as discussed in the next section.

\subsection{Accounting for Encoding Complexity}
In the following section, we delve into a complexity analysis examining the fundamental operations required by each encoding scheme.
Since the decoding complexity is negligible compared to the encoding complexity, we will focus on the latter one.
Let us define the complexity cost of the common operations within the analyzed schemes. 
In general, these costs are a function of the bit sequence length $n$ and the particular technology adopted.
We define $\mathcal{C}_\mathrm{w}(n)$ as the cost to perform an exor between two sequences of length $n$ and then compute the weight of the resulting sequence (i.e., an integer representing the number of ``1''s in the sequence). 
Note that, to encode weights of $n$ long sequences, $\lceil \log_2 n \rceil$ bits are enough.
Furthermore, we define $\mathcal{C}_\mathrm{c}(n)$ as the cost to perform a comparison between two $n$ bit sequences representing an integer value such as a ``$<$'' or a ``$\le$''. 
Lastly, let us define the cost of an addition or subtraction of two $n$ long sequences representing integer values as $\mathcal{C}_\mathrm{a}(n)$. Since comparisons can be implemented using subtractions we will assume that $\mathcal{C}_\mathrm{a}(n) = \mathcal{C}_\mathrm{c}(n)$.

For the \ac{DBI} optimal scheme it is required to perform the comparison with the previous bus sequence and two new sequences, the original sequence to transmits and its bitwise not. 
Thus, the encoding complexity is given by $\mathcal{C}_\mathrm{DBI} = 2 \left[ \mathcal{C}_\mathrm{w}(n) + \mathcal{C}_\mathrm{c}(\lceil \log_2 n \rceil) \right]$.
For the \ac{PI} scheme, let us assume for the sake of simplicity that $k/b$ is an integer number. 
As a consequence, we have that the encoding complexity is given by $\mathcal{C}_\mathrm{PI} = 2b \left[ \mathcal{C}_\mathrm{w}(k/b) + \mathcal{C}_\mathrm{c}(\lceil \log_2 k/b \rceil) \right]$.
For the \ac{PR} and \ac{RI} schemes we have to compare the previous bus sequence and $2^b$ test sequences. As a consequence, $\mathcal{C}_\mathrm{PR} = 2^b \left[ \mathcal{C}_\mathrm{w}(n) + \mathcal{C}_\mathrm{c}(\lceil \log_2 n \rceil) \right]$. 
Regarding the \ac{SI} scheme, the same complexity of the \ac{PR} holds, if we substitute the $2^b$ factor with $\min\{k, 2^b\}$.
The implementation of the general optimal scheme described in Section~\ref{subsec:Bound} is discussed in \cite{ValChi:24arxiv}, where it is shown that it requires an average encoding complexity of $(n + 2) D_\mathrm{opt}(k,b)\,\mathcal{C}_\mathrm{c}(n)+(d_\mathrm{max} + 1)\,\mathcal{C}_\mathrm{c}(n)$. 
The comparison of encoding complexity for the different schemes is summarized in Tab.~\ref{tab:complexity}. 

\input{StableFigures/tabComplexity}

\section{Numerical Results}
\label{sec:NumericalResults}

In this section, we present performance evaluations based on the average Hamming distance \( D(k,b) \), computed using the analytical expressions provided in  \eqref{eq:Dopt}, \eqref{eq:DPR}, \eqref{eq:DPRApprox}, \eqref{eq:DRI}, and \eqref{eq:DSI}. Among these, only the approximation in \eqref{eq:DSI} requires validation via simulation; the results for all other schemes are derived exactly. 
We adopt as our metric the bit-flip savings defined in \eqref{eq:1-D/Dunc}. In the context of bus encoding, this metric can be directly interpreted as energy savings when the bus capacitance is sufficiently large, such that the encoding overhead is negligible. 
Otherwise, the bit-flip saving is an upper bound on the achievable energy saving.

\begin{figure}[t]
    \centering
    \resizebox{0.99\columnwidth}{!}{
\input{Figures/EnergySaving_OptimumBound.tex}
    }
    \caption{Bit-Flip saving $1-D(k,b)/D_\mathrm{unc}(k)$ for the optimal  scheme, information size $k=11$.}
 \label{fig:optimum_k11}
\end{figure}
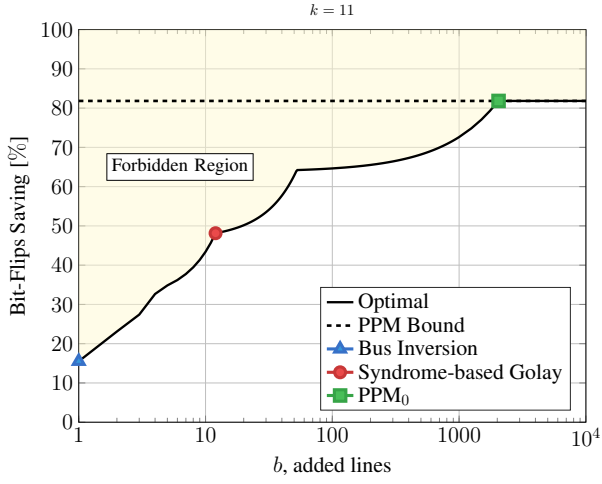

First, we show the bit-flips saving obtainable with the optimal scheme, when varying the number of added bus lines $b$. To this aim, in Fig.~\ref{fig:optimum_k11} we report the saving given by $D_\mathrm{opt}(k,b)$  derived in Sec.~\ref{subsec:Bound}, for $k=11$ information bits. By increasing $b$, the largest saving is obtained with the maximum redundancy scheme, which is the PPM where only weight-0 and weight-1 codewords are used, requiring  $b=2^k-1-k=2036$. 
It is interesting to note that for some values of $b$ the implementation of the optimal scheme is easy to understand: i) for $b=1$ the scheme is the \ac{DBI} of Sec.~\ref{subsec:Singleinv}; ii) for $b=2036$ we have the maximum redundancy scheme, i.e., \ac{PPM} plus the all-zero vector; iii) for $b=12$ we can interpret the $11$ information bits as the syndrome of the $(23,12)$ Golay code, and use a Golay decoder \cite{Eli:87} to map all configurations of the $11$ bits patterns to all patterns of length $23$ with maximum weight $3$.   

\begin{figure}[t]
 \centering
 \resizebox{0.95\columnwidth}{!}{
\input{Figures/EnergySaving_k64}
 }
 \caption{Bit-flip saving \( 1 - D(k,b)/D_\mathrm{unc}(k) \) for various encoding schemes, with information size \( k = 64 \).}
 \label{fig:perf_k64}
\end{figure}
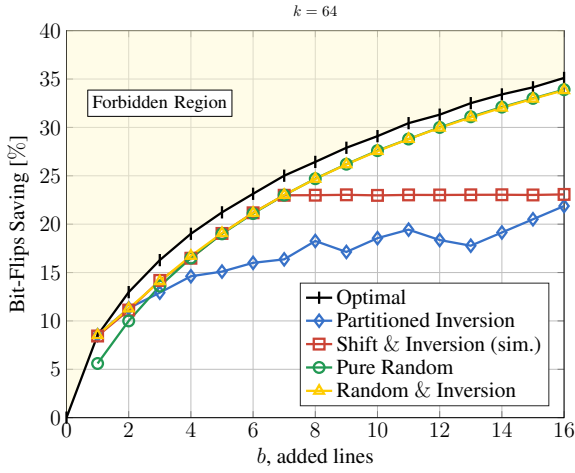

Fig.~\ref{fig:perf_k64} illustrates the bit-flip savings achieved by different encoding strategies for \( k = 64 \) bits. Notably, the simulation results for the \acl{SI} scheme closely match those of the \acl{RI} scheme, as predicted by \eqref{eq:DSI}. Up to \( b = 7 \) added lines, \acl{SI} performs nearly identically to \acl{RI}. Furthermore, for \( b > 3 \), the \acl{PR} scheme also converges in performance with the \acl{RI} approach. All of these schemes approach the performance of the optimal code with minimal degradation. For example, with \( b = 16 \) (corresponding to 25\% redundancy), the optimal scheme achieves a bit-flip saving of approximately 35\%, while both the \acl{PR} and \acl{RI} schemes attain around 34\%. In contrast, the \acl{PI} scheme yields a noticeably lower saving of about 22\%.

\begin{figure}[t]
 \centering
 \resizebox{0.95\columnwidth}{!}{
     \input{Figures/EnergySaving_k128}
 }
 \caption{Bit-flip saving \( 1 - D(k,b)/D_\mathrm{unc}(k) \) for various encoding schemes, with information size \( k = 128 \).}
 \label{fig:perf_k128}
\end{figure}
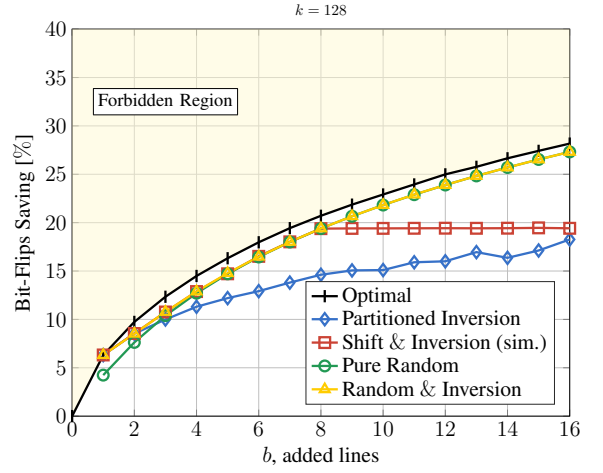

Fig.~\ref{fig:perf_k128} reports similar results for \( k = 128 \) bits. For instance, with \( b = 8 \) added bits (a 6.25\% increase in bus width), the optimal scheme yields a saving of 20.7\%, while the Pure Random and Random \& Inversion schemes reach 19.4\%. The Partitioned Inversion scheme achieves a lower saving of approximately 14.6\%.

To enable a direct numerical comparison, Table~\ref{tab:summary} summarizes selected results, where \( \eta \) denotes the bit-flip saving. 
\textcolor{black}{The table compares four coding schemes based on their configuration of information bits $k$, redundancy bits $b$, code rate $k/n$, and energy savings $\eta$. 
For each scheme, different values of $k$ and $b$ are considered to evaluate how the redundancy affects the system's energy efficiency.
In general, increasing the redundancy $b$ lowers the code rate $k/n$ but leads to higher energy savings. 
DBI uses minimal redundancy and achieves high code rates, but results in relatively lower energy savings. 
Both PI and RI introduce additional redundancy and achieve improved energy savings, with RI consistently outperforming PI and providing greater gains as the parameter $b$ increases.
}

\textcolor{black}{
While this study offers valuable insights into the potential energy savings of various bus encoding techniques, future work is necessary to assess their impact on overall system energy efficiency within specific application contexts. 
In particular, subsequent research should integrate implementation-level factors such as the energy consumption of additional encoder and decoder circuitry, physical chip area, and latency constraints. 
A comprehensive analysis that accounts for these practical considerations will be crucial to determine whether the theoretical energy savings translate into net benefits in real-world deployments.
}

\input{StableFigures/tableSummary2}

\section{Conclusions}\label{sec:conclusions}


We have developed a novel analysis to assess the bit-flips saving across various encoding schemes, facilitating the comparison among different approaches without the need for simulations. We also proposed two novel practical schemes, \acl{RI} and \acl{SI}, which together with existing previous proposals allow to obtain a substantial reduction of bit-flips for non-volatile memories and bus encoding. Our findings indicate that suitably designed sub-optimal techniques yield performance closely approximating that of the optimal, but high-complexity, solution, while contributing to more energy-efficient designs. Such improvements are particularly relevant for large-scale deployments in data centers, where reducing energy consumption and extending memory endurance are critical.

\section*{Acknowledgment}
Work supported in part by the European Union through the Italian National Recovery and Resilience Plan of NextGenerationEU, partnership on ``Telecommunications of the Future'' under Grant PE00000001-RESTART.
The authors would like to thank Prof.~Luca Benini and Prof.~Enrico Paolini for their comments and suggestions.  

\bibliographystyle{IEEEtran}
\bibliography{Files/IEEEabrv,Files/StringDefinitions,Files/StringDefinitions2,Files/refs}

\end{document}

%% file: Files/Acronimi_CD.tex
\begin{acronym}
\small
\acro{i.i.d.}{independent and identically distributed}
\acro{BSC}{binary symmetric channel}
\acro{CNN}{convolutional neural network}
\acro{ECC}{error correcting code}
\acro{LUT}{look-up-table}
\acro{i.i.d.}{independent and identically distributed}
\acro{MPPM}{multi pulse-position modulation}
\acro{MUX}{multiplexer}
\acro{NVM}{non-volatile memories}
\acro{NoC}{Network-on-Chip}
\acro{PI}{Partitioned Inversion}
\acro{PRAM}{phase-change random access memory}
\acro{PPM}{pulse-position modulation}
\acro{PR}{Pure Random}
\acro{RI}{Random Inversion}
\acro{r.v.}{random variable}
\acro{DBI}{Data Bus Inversion}
\acro{DMS}{discrete memoryless source}
\acro{SI}{Shift Inversion}
\acro{SSO}{simultaneous switching output}
\acro{WEM}{Write-Efficient Memory}
\end{acronym}

%% file: Figures/GeneralFrameworkFinal.tex
\tikzset{every picture/.style={line width=0.75pt}} 

\begin{tikzpicture}[x=0.75pt,y=0.75pt,yscale=-1,xscale=1]

\draw  [fill={rgb, 255:red, 184; green, 233; blue, 134 }  ,fill opacity=0.6 ] (100,99.5) -- (210,99.5) -- (210,119.5) -- (100,119.5) -- cycle ;
\draw   (260,80) -- (350,80) -- (350,120) -- (260,120) -- cycle ;
\draw    (220,110) -- (258,110) ;
\draw [shift={(260,110)}, rotate = 180] [color={rgb, 255:red, 0; green, 0; blue, 0 }  ][line width=0.75]    (6.56,-1.97) .. controls (4.17,-0.84) and (1.99,-0.18) .. (0,0) .. controls (1.99,0.18) and (4.17,0.84) .. (6.56,1.97)   ;
\draw    (290,50) -- (220,50) -- (220,90) -- (258,90) ;
\draw [shift={(260,90)}, rotate = 180] [color={rgb, 255:red, 0; green, 0; blue, 0 }  ][line width=0.75]    (6.56,-1.97) .. controls (4.17,-0.84) and (1.99,-0.18) .. (0,0) .. controls (1.99,0.18) and (4.17,0.84) .. (6.56,1.97)   ;
\draw    (350,100) -- (428,100) ;
\draw [shift={(430,100)}, rotate = 180] [color={rgb, 255:red, 0; green, 0; blue, 0 }  ][line width=0.75]    (6.56,-1.97) .. controls (4.17,-0.84) and (1.99,-0.18) .. (0,0) .. controls (1.99,0.18) and (4.17,0.84) .. (6.56,1.97)   ;
\draw  [fill={rgb, 255:red, 255; green, 249; blue, 210 }  ,fill opacity=1 ] (430,80) -- (530,80) -- (530,120) -- (430,120) -- cycle ;
\draw  [fill={rgb, 255:red, 254; green, 209; blue, 137 }  ,fill opacity=1 ] (320,0) -- (470,0) -- (470,20) -- (320,20) -- cycle ;
\draw    (700,110) -- (738,110) ;
\draw [shift={(740,110)}, rotate = 180] [color={rgb, 255:red, 0; green, 0; blue, 0 }  ][line width=0.75]    (6.56,-1.97) .. controls (4.17,-0.84) and (1.99,-0.18) .. (0,0) .. controls (1.99,0.18) and (4.17,0.84) .. (6.56,1.97)   ;
\draw   (290,30) -- (330,30) -- (330,70) -- (290,70) -- cycle ;
\draw    (380,100) -- (380,50) -- (332,50) ;
\draw [shift={(330,50)}, rotate = 360] [color={rgb, 255:red, 0; green, 0; blue, 0 }  ][line width=0.75]    (6.56,-1.97) .. controls (4.17,-0.84) and (1.99,-0.18) .. (0,0) .. controls (1.99,0.18) and (4.17,0.84) .. (6.56,1.97)   ;
\draw  [fill={rgb, 255:red, 0; green, 0; blue, 0 }  ,fill opacity=1 ] (377.13,100) .. controls (377.13,98.42) and (378.42,97.13) .. (380,97.13) .. controls (381.58,97.13) and (382.87,98.42) .. (382.87,100) .. controls (382.87,101.58) and (381.58,102.87) .. (380,102.87) .. controls (378.42,102.87) and (377.13,101.58) .. (377.13,100) -- cycle ;
\draw    (530,99.5) -- (608,99.99) ;
\draw [shift={(610,100)}, rotate = 180.36] [color={rgb, 255:red, 0; green, 0; blue, 0 }  ][line width=0.75]    (6.56,-1.97) .. controls (4.17,-0.84) and (1.99,-0.18) .. (0,0) .. controls (1.99,0.18) and (4.17,0.84) .. (6.56,1.97)   ;
\draw   (610,80) -- (700,80) -- (700,120) -- (610,120) -- cycle ;
\draw   (630,30) -- (670,30) -- (670,70) -- (630,70) -- cycle ;
\draw    (580.13,99.5) -- (580,50) -- (628,50) ;
\draw [shift={(630,50)}, rotate = 180] [color={rgb, 255:red, 0; green, 0; blue, 0 }  ][line width=0.75]    (6.56,-1.97) .. controls (4.17,-0.84) and (1.99,-0.18) .. (0,0) .. controls (1.99,0.18) and (4.17,0.84) .. (6.56,1.97)   ;
\draw    (670,50) -- (740,50) -- (740,90) -- (702,90) ;
\draw [shift={(700,90)}, rotate = 360] [color={rgb, 255:red, 0; green, 0; blue, 0 }  ][line width=0.75]    (6.56,-1.97) .. controls (4.17,-0.84) and (1.99,-0.18) .. (0,0) .. controls (1.99,0.18) and (4.17,0.84) .. (6.56,1.97)   ;
\draw  [fill={rgb, 255:red, 0; green, 0; blue, 0 }  ,fill opacity=1 ] (577.27,99.5) .. controls (577.27,97.92) and (578.55,96.63) .. (580.13,96.63) .. controls (581.72,96.63) and (583,97.92) .. (583,99.5) .. controls (583,101.08) and (581.72,102.37) .. (580.13,102.37) .. controls (578.55,102.37) and (577.27,101.08) .. (577.27,99.5) -- cycle ;

\draw (154,108) node   [align=left] {$\displaystyle k$ bit};
\draw (73,117.6) node [anchor=south] [inner sep=0.75pt]    {$\mathbf{u}_{j}$};
\draw (403,89.6) node [anchor=south] [inner sep=0.75pt]    {$\mathbf{x}_{j}$};
\draw (305,100) node   [align=left] {\begin{minipage}[lt]{36.7pt}\setlength\topsep{0pt}
\begin{center}
Encoder
\end{center}

\end{minipage}};
\draw (481.5,99.5) node   [align=left] {\begin{minipage}[lt]{56.54pt}\setlength\topsep{0pt}
\begin{center}
Bus/Memory
\end{center}

\end{minipage}};
\draw (239.5,80.6) node [anchor=south] [inner sep=0.75pt]    {$\mathbf{x}_{j-1}$};
\draw (658,99.5) node   [align=left] {\begin{minipage}[lt]{37.25pt}\setlength\topsep{0pt}
\begin{center}
Decoder
\end{center}

\end{minipage}};
\draw (755.5,117.6) node [anchor=south] [inner sep=0.75pt]    {$\mathbf{u}_{j}$};
\draw (719.5,80.6) node [anchor=south] [inner sep=0.75pt]    {$\mathbf{x}_{j-1}$};
\draw (558,90.6) node [anchor=south] [inner sep=0.75pt]    {$\mathbf{x}_{j}$};
\draw (310,50) node  [font=\Large]  {$T$};
\draw (395,10) node   [align=left] {$\displaystyle n=k+b$ \ bit};
\draw (650,50) node  [font=\Large]  {$T$};
\draw (306,18.6) node [anchor=south] [inner sep=0.75pt]    {$\mathbf{x}_{j}$};

\end{tikzpicture}

%% file: StableFigures/tabComplexity.tex
\begin{table}[t]
    \centering
    \caption{{Complexity analysis summary in terms of elementary operations.}}
    \label{tab:complexity}
    \small
    {
    \begin{tabular}{cl}
        \toprule
        \textbf{Scheme} & \multicolumn{1}{c}{$\mathcal{C}_\mathrm{enc}$}\\
        \midrule
        DBI   & $2 \times \left[ \mathcal{C}_\mathrm{w}(n) + \mathcal{C}_\mathrm{c}(\lceil \log_2 n \rceil) \right]$ \\
        PI  & $2b \times \left[ \mathcal{C}_\mathrm{w}(k/b) + \mathcal{C}_\mathrm{c}(\lceil \log_2 k/b \rceil) \right]$ \\
        PR/RI  & $2^b \times \left[ \mathcal{C}_\mathrm{w}(n) + \mathcal{C}_\mathrm{c}(\lceil \log_2 n \rceil) \right]$ \\
        SI  & $\min\{k,2^b\} \times \left[ \mathcal{C}_\mathrm{w}(n) + \mathcal{C}_\mathrm{c}(\lceil \log_2 n \rceil) \right]$ \\
        Gen. Opt. & $\left[(n + 2) D_\mathrm{opt}(k,b)+d_\mathrm{max} + 1\right] \times \mathcal{C}_\mathrm{c}(n)$ \\
        \bottomrule
    \end{tabular}
    }
\end{table}

%% file: Figures/EnergySaving_OptimumBound.tex
%
%
\definecolor{testGreen}{HTML}{36a447}
\definecolor{testBlue}{HTML}{3771c8}
\definecolor{testRed}{HTML}{c83737}
\definecolor{testYellow}{HTML}{ffcc00}
\definecolor{testViolet}{HTML}{834dd3}
\definecolor{testOrange}{HTML}{ea7527}

\definecolor{innerRed}{HTML}{ea4a4a}
\definecolor{innerGreen}{HTML}{48be58}

\definecolor{innerYellow}{HTML}{FFF9D2}

\begin{tikzpicture}
\begin{axis}[%
width=4.521in,
height=3.5in,
at={(0in,0in)},
scale only axis,
xmode=log,
xmin=1,
xmax=10000,
title={$k=11$},
ytick distance=0.1,
xtick = {1, 10, 100, 1e3, 1e4},
xticklabels = {$1$, $10$, $100$, $1000$, $10^4$},
xminorticks=true,
xlabel style={font=\color{white!15!black}, font=\Large},
xlabel={$b$, added lines},
ymin=0,
ymax=1,
ylabel style={font=\color{white!15!black}, font=\Large},
ytick = {0, 0.1, 0.2, 0.3, 0.4, 0.5, 0.6, 0.7, 0.8, 0.9, 1},
yticklabels={$0$, $10$, $20$, $30$, $40$, $50$, $60$, $70$, $80$, $90$, $100$},
ylabel style={font=\color{white!15!black}, font=\Large},
ylabel={Bit-Flips Saving [$\%$]},
tick label style={black, semithick, font=\Large},
axis background/.style={fill=white},
xmajorgrids,
ymajorgrids,
legend style={at={(0.97,0.03)}, anchor=south east, legend cell align=left, align=left, draw=white!15!black, font=\Large}
]
\addplot [name path=LB,color=black, line width=1.5pt]
  table[row sep=crcr]{%
0	0\\
1	0.155184659090909\\
2	0.231001420454545\\
3	0.274058948863636\\
4	0.326615767045455\\
5	0.348366477272727\\
6	0.362127130681818\\
7	0.377485795454545\\
8	0.39453125\\
9	0.413352272727273\\
10	0.434037642045455\\
11	0.456676136363636\\
12	0.481356534090909\\
13	0.483575994318182\\
14	0.485884232954545\\
15	0.48828125\\
16	0.490767045454545\\
17	0.493341619318182\\
18	0.496004971590909\\
19	0.498757102272727\\
20	0.501598011363636\\
21	0.504527698863636\\
22	0.507546164772727\\
23	0.510653409090909\\
24	0.513849431818182\\
25	0.517134232954545\\
26	0.5205078125\\
27	0.523970170454545\\
28	0.527521306818182\\
29	0.531161221590909\\
30	0.534889914772727\\
31	0.538707386363636\\
32	0.542613636363636\\
33	0.546608664772727\\
34	0.550692471590909\\
35	0.554865056818182\\
36	0.559126420454545\\
37	0.5634765625\\
38	0.567915482954545\\
39	0.572443181818182\\
40	0.577059659090909\\
41	0.581764914772727\\
42	0.586558948863636\\
43	0.591441761363636\\
44	0.596413352272727\\
45	0.601473721590909\\
46	0.606622869318182\\
47	0.611860795454545\\
48	0.6171875\\
49	0.622602982954545\\
50	0.628107244318182\\
51	0.633700284090909\\
52	0.639382102272727\\
53	0.642223011363636\\
54	0.642311789772727\\
55	0.642400568181818\\
56	0.642489346590909\\
57	0.642578125\\
58	0.642666903409091\\
59	0.642755681818182\\
60	0.642844460227273\\
61	0.642933238636364\\
62	0.643022017045455\\
63	0.643110795454545\\
64	0.643199573863636\\
65	0.643288352272727\\
66	0.643377130681818\\
67	0.643465909090909\\
68	0.6435546875\\
69	0.643643465909091\\
70	0.643732244318182\\
71	0.643821022727273\\
72	0.643909801136364\\
73	0.643998579545455\\
74	0.644087357954545\\
75	0.644176136363636\\
76	0.644264914772727\\
77	0.644353693181818\\
78	0.644442471590909\\
79	0.64453125\\
80	0.644620028409091\\
81	0.644708806818182\\
82	0.644797585227273\\
83	0.644886363636364\\
84	0.644975142045455\\
85	0.645063920454545\\
86	0.645152698863636\\
87	0.645241477272727\\
88	0.645330255681818\\
89	0.645419034090909\\
90	0.6455078125\\
91	0.645596590909091\\
92	0.645685369318182\\
93	0.645774147727273\\
94	0.645862926136364\\
95	0.645951704545455\\
96	0.646040482954545\\
97	0.646129261363636\\
98	0.646218039772727\\
99	0.646306818181818\\
100	0.646395596590909\\
125	0.648615056818182\\
150	0.650834517045455\\
175	0.653053977272727\\
200	0.6552734375\\
225	0.657492897727273\\
250	0.659712357954545\\
275	0.661931818181818\\
300	0.664151278409091\\
325	0.666370738636364\\
350	0.668590198863636\\
375	0.670809659090909\\
400	0.673029119318182\\
425	0.675248579545455\\
450	0.677468039772727\\
475	0.6796875\\
500	0.681906960227273\\
525	0.684126420454545\\
550	0.686345880681818\\
575	0.688565340909091\\
600	0.690784801136364\\
625	0.693004261363636\\
650	0.695223721590909\\
675	0.697443181818182\\
700	0.699662642045455\\
725	0.701882102272727\\
750	0.7041015625\\
775	0.706321022727273\\
800	0.708540482954545\\
825	0.710759943181818\\
850	0.712979403409091\\
875	0.715198863636364\\
900	0.717418323863636\\
925	0.719637784090909\\
950	0.721857244318182\\
975	0.724076704545455\\
1000	0.726296164772727\\
1250	0.748490767045455\\
1500	0.770685369318182\\
1750	0.792879971590909\\
2000	0.815074573863636\\
2036	0.818270596590909\\
3000    0.818270596590909\\
4000    0.818270596590909\\
5000    0.818270596590909\\
6000    0.818270596590909\\
7000    0.818270596590909\\
8000    0.818270596590909\\
9000    0.818270596590909\\
10000    0.818270596590909\\
};
\addlegendentry{Optimal}

\addplot [name path =asseX,color=black, draw opacity=0, dashed, line width=1.2pt,forget plot]
  table[row sep=crcr]{%
1	      1\\
10000     1\\
};
\addplot [thick, color=innerYellow, fill=innerYellow, fill opacity=0.5,forget plot]
fill between[
    of = LB and asseX,
    soft clip = {domain=1:1e4},
];

\addplot [color=black, dashed, line width=1.7pt]
  table[row sep=crcr]{%
1	    0.818270596590909\\
10000	0.818270596590909\\
};
\addlegendentry{PPM Bound}

\addplot [color=testBlue, line width=2pt, mark size=4pt, mark=triangle*, mark options={solid, fill=brightBlue}]
  table[row sep=crcr]{%
1	0.1552\\
};
\addlegendentry{Bus Inversion}

\addplot [color=testRed, line width=2pt, mark size=3.5pt, mark=*, mark options={solid, fill=innerRed}]
  table[row sep=crcr]{%
12	0.4814\\
};
\addlegendentry{Syndrome-based Golay}

\addplot [color=testGreen, line width=2pt, mark size=3.7pt, mark=square*, mark options={solid, fill=innerGreen}]
  table[row sep=crcr]{%
2047	0.818270596590909\\
};
\addlegendentry{PPM$_0$}

\node[right] (A) at (axis cs:1.5,0.65) {\fcolorbox{black}{white}{\large Forbidden Region}};

\end{axis}


\end{tikzpicture}%

%% file: Figures/EnergySaving_k64.tex
%
%
\definecolor{testGreen}{HTML}{36a447}
\definecolor{testBlue}{HTML}{3771c8}
\definecolor{testRed}{HTML}{c83737}
\definecolor{testYellow}{HTML}{ffcc00}
\definecolor{testViolet}{HTML}{834dd3}
\definecolor{testOrange}{HTML}{ea7527}
\definecolor{innerRed}{HTML}{ea4a4a}
\definecolor{innerGreen}{HTML}{48be58}
\definecolor{innerYellow}{HTML}{FFF9D2}
\begin{tikzpicture}

\begin{axis}[%
width=4.5in,
height=3.5in,
at={(0in,0in)},
scale only axis,
title={$k=64$},
xmin=0,
xmax=16,
ymin=0,
ymax=0.4,
xtick={ 0,  2,  4,  6,  8, 10, 12, 14, 16},
ytick = {0, 0.05, 0.1, 0.15, 0.2, 0.25, 0.3, 0.35, 0.4},
yticklabels={$0$, $5$, $10$, $15$, $20$, $25$, $30$, $35$, $40$},
xlabel style={font=\color{white!15!black}, font=\Large},
xlabel={$b$, added lines},
ylabel style={font=\color{white!15!black}, font=\Large},
ylabel={Bit-Flips Saving [$\%$]},
tick label style={black, semithick, font=\Large},
axis background/.style={fill=white},
xmajorgrids,
ymajorgrids,
legend style={at={(0.97,0.03)}, anchor=south east, legend cell align=left, align=left, draw=white!15!black, font=\Large}
]

\addplot [name path =LB,color=black, line width=1.5pt, mark=|, mark size = 4.0pt, mark options={solid, black}]
  table[row sep=crcr]{%
0	0                  \\
1	0.085274046775280  \\
2	0.129737333676810  \\
3	0.162854880732230  \\
4	0.190063656624960  \\
5	0.212220491253940  \\
6	0.231404871925700  \\
7	0.250107543212020  \\
8	0.264356009418640  \\
9	0.278962358207450  \\
10	0.290979334443690  \\
11	0.304404594045540  \\
12	0.313170452998480  \\
13	0.325060275543390  \\
14	0.334102383296050  \\
15	0.341442708386730  \\
16	0.351126471337730  \\
};
\addlegendentry{Optimal}

\addplot [name path =asseX,color=black, draw opacity=0, dashed, line width=1.2pt,forget plot]
  table[row sep=crcr]{%
0	1\\
16	1\\
};
\addplot [thick, color=innerYellow, fill=innerYellow, fill opacity=0.5,forget plot]
fill between[
    of = LB and asseX,
    soft clip = {domain=0:16},
];

\addplot [color=testBlue, line width=1.5pt, mark=diamond, mark size = 4.0pt, mark options={solid, testBlue}]
  table[row sep=crcr]{%
1	0.085280  \\
2	0.113080  \\
3	0.129200  \\
4	0.146160  \\
5	0.150990  \\
6	0.160040  \\
7	0.163640  \\
8	0.182620  \\
9	0.171270  \\
10	0.185550  \\
11	0.194340  \\
12	0.183600  \\
13	0.177740  \\
14	0.191410  \\
15	0.205080  \\
16	0.218750  \\
};
\addlegendentry{Partitioned Inversion}

\addplot [color=brightRed, line width=1.5pt, mark=square, mark size = 3.5pt, mark options={solid, brightRed}]
  table[row sep=crcr]{%
1	0.084410  \\
2	0.111220  \\
3	0.141850  \\
4	0.164790  \\
5	0.1904125\\
6	0.211747916666667\\
7	0.229863541666667\\
8	0.229946875\\
9	0.230346875\\
10	0.229746875\\
11	0.23025625\\
12	0.230173958333333\\
13	0.230334375\\
14	0.230410  \\
15	0.230190  \\
16	0.230750  \\
};
\addlegendentry{Shift $\&$ Inversion (sim.)}

\addplot [color=graphGreen, line width=1.5pt, mark=o, mark size = 3.5pt, mark options={solid, graphGreen}]
  table[row sep=crcr]{%
1	0.0560  \\
2	0.1000  \\
3	0.1360  \\
4	0.1650  \\
5	0.1900  \\
6	0.2110  \\
7	0.2300  \\
8	0.2470  \\
9	0.2620  \\
10	0.2760  \\
11	0.2880  \\
12	0.3000  \\
13	0.3110  \\
14	0.3210  \\
15	0.3300  \\
16	0.3390  \\
};
\addlegendentry{Pure Random}


\addplot [color=testYellow, line width=1.5pt, mark=triangle, mark size = 3.7pt, mark options={solid, testYellow}]
  table[row sep=crcr]{%
1	0.08527\\
2	0.1128\\
3	0.1409\\
4	0.1672\\
5	0.1907\\
6	0.2116\\
7	0.2301\\
8	0.2468\\
9	0.2618\\
10	0.2755\\
11	0.288\\
12	0.2996\\
13	0.3103\\
14	0.3203\\
15	0.3296\\
16	0.3384\\  
};
\addlegendentry{Random $\&$ Inversion}

\node[right] (A) at (axis cs:0.5,0.325) {\fcolorbox{black}{white}{\large Forbidden Region}};

\end{axis}
\end{tikzpicture}%

%% file: Figures/EnergySaving_k128.tex
%
%
\definecolor{testGreen}{HTML}{36a447}
\definecolor{testBlue}{HTML}{3771c8}
\definecolor{testRed}{HTML}{c83737}
\definecolor{testYellow}{HTML}{ffcc00}
\definecolor{testViolet}{HTML}{834dd3}
\definecolor{testOrange}{HTML}{ea7527}

\definecolor{innerRed}{HTML}{ea4a4a}
\definecolor{innerGreen}{HTML}{48be58}

\definecolor{innerYellow}{HTML}{FFF9D2}
\begin{tikzpicture}

\begin{axis}[%
width=4.5in,
height=3.5in,
at={(0in,0in)},
scale only axis,
title={$k=128$},
xmin=0,
xmax=16,
ymin=0,
ymax=0.4,
xtick={ 0,  2,  4,  6,  8, 10, 12, 14, 16},
ytick = {0, 0.05, 0.1, 0.15, 0.2, 0.25, 0.3, 0.35, 0.4},
yticklabels={$0$, $5$, $10$, $15$, $20$, $25$, $30$, $35$, $40$},
xlabel style={font=\color{white!15!black}, font=\Large},
xlabel={$b$, added lines},
ylabel style={font=\color{white!15!black}, font=\Large},
ylabel={Bit-Flips Saving [$\%$]},
tick label style={black, semithick, font=\Large},
axis background/.style={fill=white},
xmajorgrids,
ymajorgrids,
legend style={at={(0.97,0.03)}, anchor=south east, legend cell align=left, align=left, draw=white!15!black, font=\Large}
]

\addplot [name path =LB, color=black, line width=1.5pt, mark=|, mark size = 4.0pt, mark options={solid, black}]
  table[row sep=crcr]{%
0	0 \\
1	0.06310 \\
2	0.09740 \\
3	0.12340 \\
4	0.14490 \\
5	0.16320 \\
6	0.17970 \\
7	0.19430 \\
8	0.20710 \\
9	0.21860 \\
10	0.22910 \\
11	0.23950 \\
12	0.24990 \\
13	0.25770 \\
14	0.26650 \\
15	0.27420 \\
16	0.28170 \\
};
\addlegendentry{Optimal}

\addplot [name path =asseX,color=black, draw opacity=0, dashed, line width=1.2pt,forget plot]
  table[row sep=crcr]{%
0	1\\
16	1\\
};
\addplot [thick, color=innerYellow, fill=innerYellow, fill opacity=0.5,forget plot]
fill between[
    of = LB and asseX,
    soft clip = {domain=0:16},
];

\addplot [color=testBlue, line width=1.5pt, mark=diamond, mark size = 4.0pt, mark options={solid, testBlue}]
  table[row sep=crcr]{%
1	0.063130 \\
2	0.085280 \\
3	0.099910 \\
4	0.113080 \\
5	0.121980 \\
6	0.129200 \\
7	0.138030 \\
8	0.146160 \\
9	0.150620 \\
10	0.150990 \\
11	0.159040 \\
12	0.160040 \\
13	0.169530 \\
14	0.163640 \\
15	0.171210 \\
16	0.182620 \\
};
\addlegendentry{Partitioned Inversion}

\addplot [color=brightRed, line width=1.5pt, mark=square, mark size = 3.5pt, mark options={solid, brightRed}]
  table[row sep=crcr]{%
1	0.0632338541666655\\
2	0.0854255208333322\\
3	0.107633333333332\\
4	0.128766666666666\\
5	0.147128645833332\\
6	0.165161458333332\\
7	0.180141666666666\\
8	0.193805729166666\\
9	0.194045312499999\\
10	0.194080208333332\\
11	0.194165104166666\\
12	0.194259895833332\\
13	0.194160041666666\\
14	0.194259895833332\\
15	0.194601041666666\\
16	0.194160041666666\\
};
\addlegendentry{Shift $\&$ Inversion (sim.)}

 \addplot [color=graphGreen, line width=1.5pt, mark=o, mark size = 3.5pt, mark options={solid, graphGreen}]
   table[row sep=crcr]{%
1	0.0424\\
2	0.07621\\
3	0.1039\\
4	0.1271\\
5	0.1471\\
6	0.1645\\
7	0.18\\
8	0.1939\\
9	0.2066\\
10	0.2182\\
11	0.2289\\
12	0.2389\\
13	0.2483\\
14	0.257\\
15	0.2653\\
16	0.2731\\
 };
 \addlegendentry{Pure Random}


\addplot [color=testYellow, line width=1.5pt, mark=triangle, mark size = 3.7pt, mark options={solid, testYellow}]
  table[row sep=crcr]{%
1	0.06312\\
2	0.08518\\
3	0.1077\\
4	0.1288\\
5	0.1478\\
6	0.1648\\
7	0.1801\\
8	0.194\\
9	0.2066\\
10	0.2182\\
11	0.229\\
12	0.2389\\
13	0.2483\\
14	0.257\\
15	0.2653\\
16	0.2731\\
};
\addlegendentry{Random $\&$ Inversion}

\node[right] (A) at (axis cs:0.5,0.325) {\fcolorbox{black}{white}{\large Forbidden Region}};

\end{axis}
\end{tikzpicture}%

%% file: StableFigures/tableSummary2.tex
\begin{table}[t]
    \centering
    \caption{Examples for some choices of the system parameters}
    \label{tab:summary}
    \small
    {
    \begin{tabular}{ccccccc}
        \toprule
        \textbf{Scheme} & $k$ & $D_\mathrm{unc}(k)$ & $b$ & $k/n$ & $ D(k,b)$ & $\eta$ [$\%$]\\
        \midrule
        \multirow{2}{*}{DBI} & $8$ & $4$ & $1$ & $0.889$ & $3.27$ & $18.3$\\
        \cmidrule{2-7}
         & $64$ & $32$ & $1$ & $0.985$ & $29.27$ & $8.5$\\
        \midrule
        \multirow{3}{*}{PI} & $8$ & $4$ & $2$ & $0.800$ & $3.12$ & $21.9$ \\
        \cmidrule{2-7}
        & \multirow{2}{*}{$64$} & \multirow{2}{*}{$32$} & $4$ & $0.941$ & $27.32$ & $14.6$\\
        \cmidrule{4-7}
        & & & $8$ & $0.889$ & $26.16$ & $18.3$\\
        \midrule
        \multirow{3}{*}{RI} & $8$ & $4$ & $2$ & $0.800$  & $3.14$ & $21.4$ \\
        \cmidrule{2-7}
        & \multirow{2}{*}{$64$} & \multirow{2}{*}{$32$} & $4$ & $0.941$ & $26.64$ & $16.7$\\
        \cmidrule{4-7}
        & & & $8$ & $0.889$ & $24.10$ & $24.7$ \\
        \midrule
        \multirow{3}{*}{Opt.} & $8$ & $4$ & $2$ & $0.800$  & $3.04$ & $23.8$ \\
        \cmidrule{2-7}
        & \multirow{2}{*}{$64$} & \multirow{2}{*}{$32$} & $4$ & $0.941$ & $25.92$ & $19.0$\\
        \cmidrule{4-7}
        & & & $8$ & $0.889$ & $23.54$ & $26.4$ \\
        \bottomrule
    \end{tabular}
    }
\end{table}